\documentclass[a4paper,12pt]{article}
\usepackage{jcappub} 
\usepackage{cite}
\usepackage{graphicx}
\usepackage{enumerate}
\usepackage{cancel}
\usepackage{color}
\usepackage{graphicx}
\usepackage{amsmath}
\usepackage{amsfonts}
\usepackage{amssymb}
\usepackage{rotating}
\usepackage{booktabs}
\usepackage{xcolor}
\usepackage{soul}

\def\comment#1{}

\title{\boldmath 
$W$ boson mass tension caused by its right-handed gauge coupling 
at high energies?}

\author{She-Sheng Xue}

\affiliation{ICRANet Piazzale della Repubblica, 10 -65122, Pescara, Italy, \\ Physics Department, University of Rome La Sapienza, \\ P.le Aldo Moro 5, I–00185 Rome, Italy\\
INFN, Sezione di Perugia, Via A. Pascoli, I-06123, Perugia, Italy
}

\emailAdd{xue@icra.it and shesheng.xue@gmail.com} 

\abstract{The CDF collaboration's recent high-precision measurement of the $W$mass is in $7.0~\sigma$ disagreement with the Standard Model expectation.
This tension will be relieved if the $W$ boson has a non-trivial right-handed gauge coupling at high energies.
At TeV scales, the SM gauge symmetric four-fermion interactions induce a right-handed gauge coupling, and SM fermions compose massive composite particles. We investigate the top-quark mass produced by spontaneous symmetry breaking and compute the $W$ and $Z$ boson propagators and decays.
The right-handed coupling corrections to their masses and widths are consistent with experimental measurements.
We discuss how SM gauge bosons and composite particles can restore parity-preserving gauge symmetries at TeV scales.      
}

\begin{document}
\maketitle
\flushbottom

%
\section
{\bf Introduction}\label{int}

The parity-violating gauge symmetries and spontaneous/explicit breaking of these symmetries for the $W$ and $Z$ gauge boson 
masses and hierarchy pattern of fermion masses have been at the 
centre of a conceptual elaboration that has played a major
role in donating to Mankind the beauty of the Standard
Model (SM) and possible scenarios beyond SM for fundamental particle physics. 
A simple description is provided on the one hand by the composite Higgs-boson model
or the Nambu-Jona-Lasinio (NJL) model \cite{Nambu:1961fr} with effective four-fermion operators, and on the other by the phenomenological model of the elementary Higgs 
boson \cite{Englert1964,Higgs1964,Guralnik1964}. 
These two models are effectively equivalent for the SM at
low energies. The ATLAS \cite{Aad2012} 
and CMS \cite{Collaboration2012}
collaborations have shown the first observations of a
125 GeV scalar particle in the search for the SM Higgs
boson. 
The interpretation of the discovered new state as the
SM Higgs boson implies that there is no anymore unknown parameter 
in the prediction for the $W$ boson mass. It is completely determined 
based on the internal symmetries of the theory and a
set of high-precision measurements of observables,
including the $Z$ and Higgs boson masses, the top-quark mass, etc.
The recent high-precision measurement 
of the $W$ boson mass \cite{Aaltonen2022} shows a difference
with a significance of $7.0 ~\sigma$ level in comparison with the 
SM expectation, including the full SM high-order 
corrections \cite{Heinemeyer2006,Heinemeyer2013}. 
This tension suggests the possibility of 
theoretical extensions to the SM.  

Several hypotheses with the additional symmetries, interactions and fields have been promulgated to provide a deeper explanation of the Higgs field, its potential, and the Higgs boson. These include
supersymmetry models \cite{Feng2013} 
and nonsupersymmetric models, e.g., compositeness, 
in which additional strong confining interactions produce
the Higgs boson as a bound state \cite{Contino2006}. 
These extensions to the SM would modify \cite{Heinemeyer2013,LopezVal2014,
Curtin2015,Chakrabortty2012,Bellazzini2014,Pomarol2014,Giudice2007,Ge2016} 
the estimated mass of the $W$ boson relative to the SM
expectation \cite{ParticleDataGroup:2020ssz}.

To address the $W$-boson mass tension, we investigate the SM gauge symmetric four-fermion interaction of NJL-type. It not only renders the effective parity-violating SM at its infrared (IR) fixed point of $v=246$ GeV but also 
a parity-preserving theory of composite particles at its ultraviolet (UV) 
fixed point of $\Lambda\sim$ TeV scales. The $W$ boson gauge coupling is 
no longer purely left-handed but has a non-trivial right-handed coupling from 
the four-fermion interaction. As a result, 
the $W$ boson mass receives a
positive correction to its SM expectation, and the tension is relieved. 
We made an early attempt to 
study this issue \cite{Xue1990,Preparata1994}. 

In Sec.~\ref{uv}, we briefly recall the 
four-fermion interaction at a natural 
cutoff and its UV fixed point of strong coupling for 
composite particles. In Secs.~\ref{top} and \ref{tope}, 
we describe its IR fixed point of weak coupling for the SM via top-quark mass generation by spontaneous gauge symmetry breaking.
We present in Sec.~\ref{ESBSw} the discussion on
the right-handed gauge couplings at high energies. Their corrections to 
the masses and widths of the $W$ and $Z$ bosons and comparisons with experimental measurements are in Secs.~\ref{wzmass} and \ref{wzwidth}. 
The examining parity-symmetry restoration by the left and right asymmetry 
is emphasised in the last section.

\comment{
The most recent W mass 80.433±0.009 GeV \cite{Aaltonen2022} and decay width 2.085±0.042 GeV (exp), reference Tanabashi, M.; et al. (Particle Data Group) (2018). "Review of Particle Physics". Physical Review D. 98 (3): 030001. Bibcode:2018PhRvD..98c0001T. doi:10.1103/PhysRevD.98.030001
wikipeadia of w boson mass: Before 2022, measurements of the W boson mass appeared to be consistent with the Standard Model. For example, in 2021, experimental measurements of the W boson mass were assessed to converge around 80,379 ± 12 MeV.[12]
However, in April 2022, a new analysis of data that was obtained by the Fermilab Tevatron collider before its closure in 2011 determined the mass of the W boson to be 80,433 ± 9 MeV, which is seven standard deviations above that predicted by the Standard Model, meaning that if the model is correct[13] there should only be a one-trillionth chance that such a large mass would arise by non-systematic observational error.[14]
}

\section{Theoretical ultraviolet completion}\label{uv}

A well-defined quantum field theory for the SM Lagrangian requires a natural regularisation at the UV cutoff $\Lambda_{\rm cut}$ fully preserving the SM 
chiral-gauge symmetry. The UV cutoff could be the Planck or the grand unified theory scale. Quantum gravity or another new physics naturally provides such regularisation. However, the No-Go theorem \cite{Nielsen:1981xu,1981PhLB..105..219N} demonstrates the existence of right-handed neutrinos and 
the lack of consistent regularisation for the SM bilinear fermion Lagrangian to precisely preserve the SM chiral-gauge symmetries. It implies four-fermion operators for SM fermions and right-handed neutrinos at the UV cutoff. 
As a theoretical model, we adopt the four-fermion operators of the torsion-free Einstein-Cartan Lagrangian with all SM fermions and three right-handed neutrinos \cite{Xue2015, Xue:2016dpl, Xue2017}. Among four-fermion operators, we consider here one for the third quark family
\begin{eqnarray}
G_{\rm cut}\Big[(\bar\psi^{ia}_Lt_{Ra})(\bar t^b_{R}\psi_{Lib}) + (\bar\psi^{ia}_Lb_{Ra})(\bar b^b_{R}\psi_{Lib})\Big],
\label{bhl}
\end{eqnarray}
where $a$ and $b$ are the colour indexes 
of the top and bottom quarks. The $SU_L(2)$ singlets $\psi_R^a 
= t^{a}_R, b^{a}_R$ and doublet $\psi^{ia}_L=(t^{a}_L,b^{a}_L)$. 
are the eigenstates of SM electroweak interactions. 
The effective four-fermion coupling 
$G_{\rm cut}\propto {\mathcal O}(\Lambda_{\rm cut}^{-2})$ and the dimensionless coupling $G_{\rm cut}\Lambda_{\rm cut}^{2}$ depends on the running scale $\mu$.

Apart from what is possible new physics at the UV cutoff $\Lambda_{\rm cut}$ 
explaining the origin of these effective four-fermion operators (\ref{bhl}), 
it is essential to study the following aspects of the interaction and ground state. (i) What the dynamics and ground state of these operators undergo in terms of their couplings as functions of running energy scale $\mu$; (ii) Associating to these dynamics and ground states where the IR or UV stable fixed point of physical couplings locates; (iii) In the domains (scaling regions) of these stable fixed points, which physically relevant operators 
become effectively dimensional-4 ($d=4$) renormalizable 
operators following renormalization group (RG) equations (i.e. scaling laws); (iv) Which ($d>4$) 
irrelevant operators though suppressed by the UV cutoff 
scale have corrections to the relevant operators.

In the strong coupling $G_{\rm cut}\Lambda_{\rm cut}^{2}\gg 1$, 
it is a symmetric phase (strong-coupling ground state) 
where massive composite bosons and fermions 
form \cite{Xue1996,Xue2000,Xue2017}\footnote{Composite particles 
were discussed for the $SU(5)$ chiral gauge symmetry theory \cite{Eichten1986, Creutz1997}.}
\begin{eqnarray}
\Phi = Z_\Phi(\bar\psi_L \psi_R);\quad \Psi_{L,R}= Z_\Psi\Phi\psi_{L,R},
\label{comp}
\end{eqnarray} 
where the colour index sums and $SU_L(2)$ isospin index is omitted. The $Z_\Phi$ and $Z_\Psi$ are energy-dependent form factors (wave function renormalisations).
As long as their form factors do not vanish, composite particles behave as elementary particles after wave function renormalisations. An effective field theory for composite 
particles (\ref{comp}) at the energy scale $\Lambda < \Lambda_{\rm cut}$\footnote{In some previous publications, the transition energy scale is indicated by ${\mathcal E}$, while $\Lambda$ stands for the UV cutoff $\Lambda_{\rm cut}$.}
is realised in the scaling domain of the stable UV fixed point $G^c_{\rm cut}\Lambda_{\rm cut}^{2}$, which is the critical coupling of the second-order phase transition 
from the strong-coupling symmetric phase to the weak-coupling symmetry breaking phase \cite{Xue:1999xa, Xue:2014opa, Xue2017}. 
The phenomenology at the LHC of these composite particles has been 
initiated \cite{Leonardi:2018jzn}.

We point out that the effective field theory is SM gauge symmetric because composite particles carry SM quantum numbers and couple to the SM gauge bosons. They are massive $M_\Phi\Phi^\dagger\Phi$ and $M_\Psi(\bar\Psi_L\psi_R+\bar\psi_L\Psi_R)$, but exactly preserving the SM chiral (parity-violating) gauge symmetries. For example, the $W$ boson couples not only to the left-handed field 
$\psi_L$, but also to the right-hand composite field $\Psi_{R}\propto \Phi^\dagger\psi_{R} \propto (\bar\psi_R \psi_L)\psi_{R}$. Namely, the $W$ boson of the chiral $SU_L(2)$ gauge symmetry has a vectorlike (parity-preserving) coupling to 
composite fermions. It implies the parity symmetry restoration at the scale 
$\Lambda$ \cite{Xue2003}.  

When the decreasing energy scale $\mu$ is smaller than 
the energy scale $\Lambda$, the coupling $G_{\rm cut}\Lambda_{\rm cut}^{2} 
< G^c_{\rm cut}\Lambda_{\rm cut}^{2}$ runs into the weak-coupling 
phase. Via the contact interactions $(\bar\psi_L \psi_R)\Phi$ and 
$[(\bar\psi_L \psi_R)\psi_{L, R}]\Psi_{L, R}$, the composite bosons $\Phi$ and fermions $\Psi_{L, R}$ dissolve into SM fermions, as their negative binding energies ${\mathcal B}$, form factors $Z_\Phi$ and $Z_\Psi$ vanish. The dissolving dynamics is similar to composite particles (poles) dissolving into their constituents (cuts) in the energy-momentum plane, e.g.~ deuteron dissolves into a proton, and a 
neutron \cite{Weinberg1963, Weinberg1963a, Weinberg1964, Weinberg1965}.
The four-fermion interacting dynamics run into the weak-coupling phase (SM ground state), the effective operators of elementary SM fermions at the energy 
scale $\Lambda$ are
\begin{eqnarray}
G\Big[(\bar\psi^{ia}_Lt_{Ra})(\bar t^b_{R}\psi_{Lib}) + (\bar\psi^{ia}_Lb_{Ra})(\bar b^b_{R}\psi_{Lib})\Big],
\label{bhlx}
\end{eqnarray}
and the four-fermion coupling $G\propto {\mathcal O}(\Lambda^{-2})$ 
at the scale $\Lambda$. The spontaneous SM symmetry breaking (SSB) dynamics proceeds. It is shown \cite{Preparata1996, Xue2013c} that for an energetically favourable SSB ground state of the least numbers of Goldstone bosons, only one massive quark $t$ and composite Higgs boson $\bar t t$ realise. Therefore, only the first term in Eq.~(\ref{bhlx}) accounts for the SSB dynamics. It gives rise to the Bardeen, C. Hill and Lindner (BHL) top-quark condensate model \cite{Bardeen1990, Cvetic1999}, where the weak-coupling IR 
fixed point $G_c\Lambda^2=8\pi^2/N_c$ (\ref{tmassgap}) realises an effective SM theory 
of the massive top quark, composite Higgs, $W^\pm$ and $Z^0$ at the electroweak 
scale $v < \Lambda$.
The approach has been generalised to the strongly-coupled Fermi liquid for the Bose-Einstein condensate \cite{Kleinert2018} and the right-handed neutrino sector for discussing dark matter particles \cite{Xue2022a}.

In summary, the ultraviolet completion (\ref{bhl}) possesses: (i) the UV fixed point for an SM gauge symmetric theory of composite particles at the scale $\Lambda$; (ii) the IR fixed point for an SM gauge symmetry breaking theory of elementary particles at the electroweak scale $v$.

\section{Top-quark 
channel 
and effective SM Lagrangian}\label{top}

In the IR fixed point domain 
$G\Lambda^2\gtrsim G_c\Lambda^2$ 
for the SSB dynamics, we use the BHL approach \cite{Bardeen1990} to study  
the top-quark mass $m_t=-(G/2)\langle \bar t t \rangle$, $W$ and 
$Z$ masses generations, and the effective SM Lagrangian at $v$. 
The mass gap-equation reads
\begin{eqnarray}
1-\left(\frac{g^c_{t0}}{g_{t0}}\right)&=& \left(\frac{m_t}{\Lambda}\right)^2\ln\left(\frac{\Lambda}{m_t}\right)^2,
\label{tmassgap}
\end{eqnarray}
where $g_{t0}\equiv G\Lambda^2$, $g_{t0}>g^c_{t0}$ and critical value 
$g^c_{t0} =G_c\Lambda^2= 8\pi^2/N_c$ 
(the colour number $N_c=3$).   
It appears the composite Higgs scalar 
$\langle\bar tt(x)\rangle$ and  
Nambu-Goldstone bosons, i.e.~$\langle\bar t(x)\gamma_5t(x)\rangle$, $\langle\bar b(x)\gamma_5t(x)\rangle$ and $\langle\bar t(x)\gamma_5b(x)\rangle$. The latter becomes the longitudinal modes of the massive $Z^0$ and 
$W^\pm$ gauge bosons. 

The effective SM Lagrangian with the {\it bilinear} top quark kinetic term and Yukawa coupling to the composite Higgs boson $H$ at the low-energy 
scale $\mu$ is given by 
\begin{eqnarray}
L &=&  L_{\rm kinetic} + \bar g_{t}(\bar \psi_L t_RH+ {\rm h.c.})
+ \Delta L_{\rm gauge} + \Delta L_{\rm irr},
\nonumber\\ 
&+& |D_\mu H|^2-m_{_H}^2H^\dagger H
-\frac{\bar\lambda}{2}(H^\dagger H)^2.
\label{eff}
\end{eqnarray}
The renormalized Yukawa coupling $\bar g_{t}$, Higgs mass $m_{_H}$ and quartic coupling $\bar \lambda$ represent $d=4$ relevant operators in the IR scaling domain.
\comment{
\begin{eqnarray}
L &=&  L_{\rm kinetic} + g_{t0}(\bar \psi_L t_RH+ {\rm h.c.}) + \Delta L_{\rm gauge} + \Delta L_{\rm irr}
\nonumber\\ 
&+& Z_H|D_\mu H|^2-m_{_H}^2H^\dagger H
-\frac{\lambda_0}{2}(H^\dagger H)^2,
\label{eff} 
\end{eqnarray}
where the bare Yukawa coupling $g_{t0}$, static Higgs mass $m_0\approx \Lambda$ and quartic coupling $\lambda_0$ at the cutoff scale $\Lambda$.
}
The $\Delta L_{\rm gauge}$ and $L_{\rm kinetic}$ are the usual SM renormalized Lagrangians of gauge bosons, top and bottom quarks. All renormalized quantities receive 
fermion-loop contributions and define for the low-energy scale $\mu$. We add the $\Delta L_{\rm irr}$ to represent the $d>4$ irrelevant operators suppressed by at least $(v/\Lambda)$. 

The conventional renormalization $Z_\psi=1$ for fundamental 
fermions and the unconventional wave-function renormalization (form factor)
$\tilde Z_H$ for the composite Higgs boson are 
adopted
\begin{equation}
\tilde Z_{H}(\mu)=\frac{1}{\bar g^2_t(\mu)},\, \bar g_t(\mu)=\frac{Z_{HY}}{Z_H^{1/2}}g_{t0}; \quad \tilde \lambda(\mu)=\frac{\bar\lambda(\mu)}{\bar g^4_t(\mu)},\,\bar\lambda(\mu)=\frac{Z_{4H}}{Z_H^2}\lambda_0,
\label{boun0}
\end{equation}
where $Z_{HY}$ and $Z_{4H}$ are proper renormalization constants of 
the Yukawa and quartic couplings in the Lagrangian (\ref{eff}). 
The full one-loop RG equations for running couplings $\bar g_t(\mu^2)$ and $\bar \lambda(\mu^2)$ read 
\begin{eqnarray}
16\pi^2\frac{d\bar g_t}{dt} &=&\left(\frac{9}{2}\bar g_t^2-8 \bar g^2_3 - \frac{9}{4}\bar g^2_2 -\frac{17}{12}\bar g^2_1 \right)\bar g_t,
\label{reg1}\\
16\pi^2\frac{d\bar \lambda}{dt} &=&12\left[\bar\lambda^2+(\bar g_t^2-A)\bar\lambda + B -\bar g^4_t \right],\quad t=\ln\mu \label{reg2}
\end{eqnarray}
where one can find $A$, $B$ and RG equations for 
SM $SU_c(3)\times SU_L(2)\times U_Y(1)$ running  renormalised gauge couplings $\bar g^2_{1,2,3}(\mu^2)$ in Eqs.~(4.7), (4.8) of 
Ref.~\cite{Bardeen1990}. 

The SSB-generated top-quark mass gives $m_t(\mu)=\bar g_t^2(\mu)v/\sqrt{2}$. 
The pole-mass 
$m^2_H(\mu)=2\bar \lambda(\mu) v^2$, form-factor $\tilde Z_H(\mu)=1/\bar g_t^2(\mu)$ and effective quartic coupling $\tilde\lambda(\mu)$ describe a composite Higgs-boson, provided that 
$\tilde Z_H(\mu)>0$ and $\tilde\lambda(\mu)>0$ are obeyed.

\begin{figure*}[t]
\hskip0.9cm\includegraphics[height=3.40in]{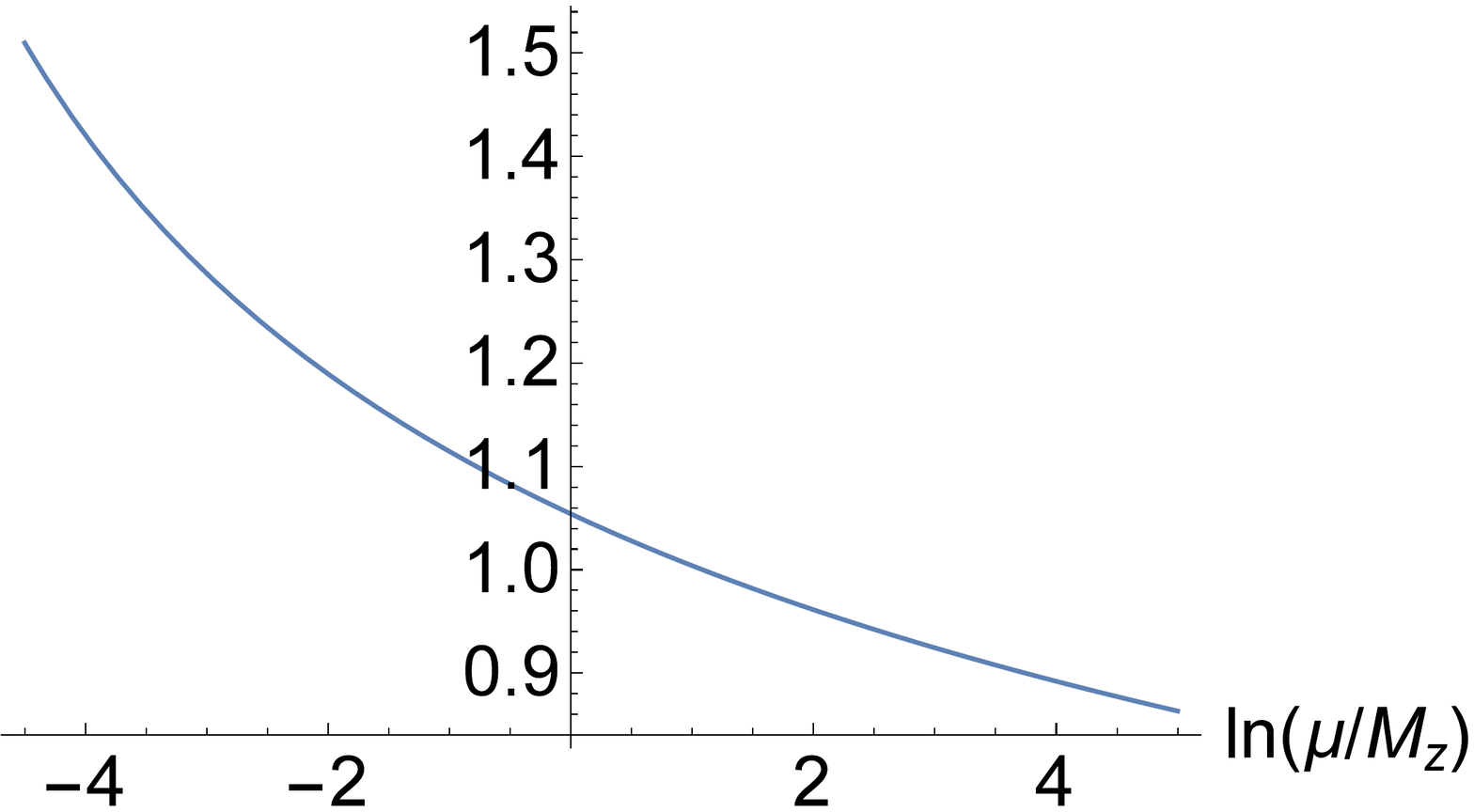}\hskip1.6cm\includegraphics[height=3.40in]{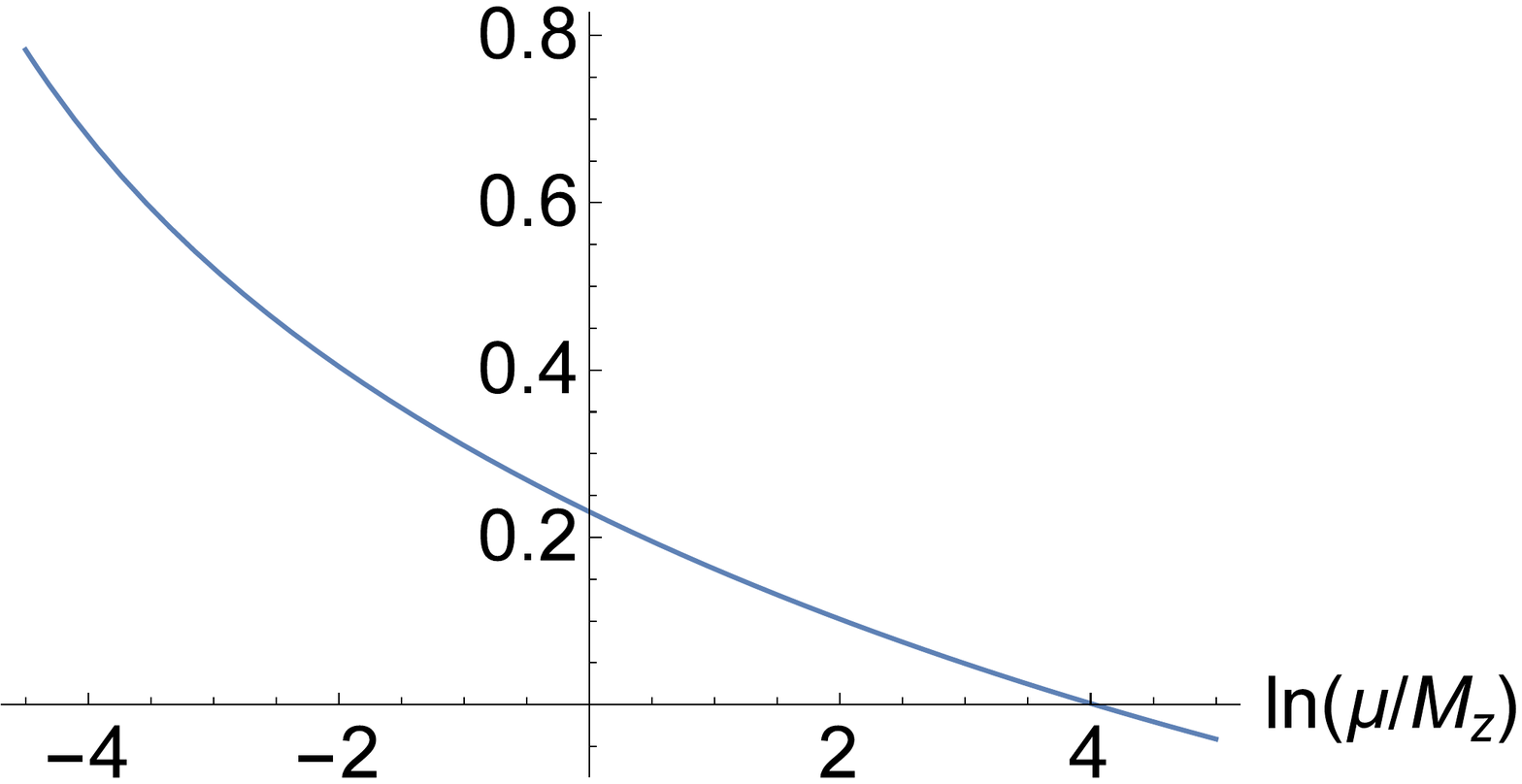}
\put(-115,175){$\tilde\lambda(\mu)$}
\put(-305,175){$\bar g_t(\mu)$}
\vskip-1.1in
\caption{In the top-quark channel, the effective (renormalised) 
Higgs Yukawa coupling $\bar g_t(\mu)$ (form factor 
$\tilde Z_H(\mu)=1/\bar g_t^2(\mu)$) and quartic coupling 
$\tilde\lambda(\mu)$ as functions of energy scale $\mu$ are determined by RG equations (\ref{reg1},\ref{reg2}), 
mass shell condition  (\ref{thshell}) of the experimentally measured top quark and Higgs mass. The effective Higgs quartic coupling $\tilde\lambda(\mu)$ becomes negative at the energy scale $\approx 5.1$ TeV.
These figures are reproduced from Refs.~\cite{xue2013,Xue:2014opa}.} 
\label{figy}
\end{figure*}

\section{Experimental measurements vs BHL composite conditions}\label{tope}

To definitely solve the RG equations (\ref{reg1}) and (\ref{reg2}) for 
$\bar g_t$ and $\bar\lambda$, one needs the boundary conditions at a physical energy scale. BHL naturally introduced the theoretical compositeness conditions,
\begin{eqnarray}
\tilde Z_H(\Lambda_{\rm cut})=1/\bar g_t^2(\Lambda_{\rm cut})=0;\quad
\tilde\lambda(\Lambda_{\rm cut})=0,
\label{bhlc}
\end{eqnarray}
at the composite scale $\sim\Lambda_{\rm cut}$, where the effective Lagrangian (\ref{eff}) is sewed together with the underlying four-fermion Lagrangian (\ref{bhl}). It is the UV completion of the BHL approach at the cutoff $\Lambda_{\rm cut}$. However, their solutions 
cannot reproduce simultaneously correct experimental values of the electroweak scale $v$, the top-quark mass $m_t$, and the Higgs boson mass 
$m_{_H}$. 

Instead, we obtained \cite{xue2013,xue2014} the solution to the RG equations (\ref{reg1}) and (\ref{reg2}) by using the boundary conditions based on the experimental values of top-quark and Higgs-boson masses, $m_t\approx 173$ GeV and $m_{_H}\approx 126$ GeV, via
the mass-shell conditions 
\begin{eqnarray}
m_t(m_t)=\bar g_t^2(m_t)v/\sqrt{2}\approx 173 {\rm GeV},
\quad m_{_H}(m_{_H})=[2\bar \lambda(m_{_H})]^{1/2} v\approx 126 {\rm GeV},
\label{thshell}
\end{eqnarray}
as well as the electroweak scale $v= 246$ GeV determined by the measurement of the Fermi constant $G_F$. 
As a result, we find the solutions for $\tilde Z_H(\mu)$ 
and $\bar\lambda(\mu)$, as shown in Fig.~\ref{figy}. In low energies $\mu\gtrsim M_z$, the effective Lagrangian and 
RG equations (\ref{eff}-\ref{reg2}) 
with experimental boundary conditions (\ref{thshell}) are equivalent 
to the SM Lagrangian 
and RG equations of elementary top-quark and Higgs fields. Extrapolating solutions 
to high energies $\mu\gg M_z$, we find that (i) $\tilde Z_H(\mu)\not=0$ is finite, the composite Higgs boson is a tightly bound state 
and behaves as an elementary particle after the 
wave-function renormalisation $\tilde Z_H(\mu)$;
(ii) the effective quartic coupling $\tilde\lambda(\mu)$ becomes negative at the energy scale $5.1$ TeV. These extrapolating results imply the aforementioned new physics beyond SM that a composite boson 
$\Phi\propto Z_\Phi(\bar t_L t_R)$ (\ref{comp}) should exist at the energy scale 
$\Lambda\approx 5.1$ TeV approximately. 

Following the BHL approach to the leading order, we obtain the decay constants of the charged and neutral Goldstone bosons,  
\begin{eqnarray}
\bar f^2(0)\approx f^2(0) \approx \frac{1}{2}N_c(4\pi)^{-2}m_t^2\ln (\Lambda^2/m_t^2)
\label{fzero}
\end{eqnarray}
at zero-momentum transfer ($q^2=0$). 
They will be explained by 
Eq.~(\ref{Wmassq}) in due course. From the tree-level $W$-boson mass 
$M_W=(1/2)\bar g_{2}(0)v$, the gauge-boson decay constant yields
\begin{eqnarray}
\frac{G_F}{\sqrt{2}} = \frac{1}{8\bar f^2(0)};\quad \bar f(0)=\frac{v}{2}.
\label{vev}
\end{eqnarray} 
On the other hand, from Eqs.~(\ref{fzero}) and (\ref{vev}), 
the gap equation (\ref{tmassgap}) becomes,
\begin{eqnarray}
1-\left(\frac{g^c_{t0}}{g_{t0}}\right)&=& \frac{(4\pi)^2}{2N_c}\left(\frac{v}{\Lambda}\right)^2.
\label{vgap}
\end{eqnarray}
These results are consistent with those given in 
Ref.~\cite{Bardeen1990} for the SM at low energies, apart from
the energy scale $\Lambda$ being much smaller than the cutoff  
$\Lambda_{\rm cut}$ of the four-fermion interaction (\ref{bhl}). 
The energy scale $\Lambda$ is an order of magnitude larger 
than the electroweak scale $v$. Therefore, 
in Eq.~(\ref{tmassgap}) or (\ref{vgap})
we do not need the fine-tuning 
$g_{t0}\gtrsim g^c_{t0}+{\mathcal O}(v^2/\Lambda^2)$ to 
achieve the electroweak scale $v$, 
While, in the BHL model of the scale $\Lambda_{\rm cut}$, 
the extremely fine-tuning $g_{t0}\gtrsim g^c_{t0}$ is required to 
remove $\Lambda_{\rm cut}^2$ and achieve $v^2\ll \Lambda^2_{\rm cut}$. 
It is also the reason why one introduces super-symmetry theories.
 
It is necessary to explain our solution in contrast with the BHL
solution. The similarities between both solutions are that they approach the scaling domain of IR fixed point at low energies $\mu\gtrsim M_z$, 
$\tilde Z_H(\mu)\not=0$ and $\tilde\lambda(\mu)\not=0$ are finite, 
the composite Higgs boson behaves as an elementary and interacting particle. 
The main differences are at high energies $\mu\gg M_z$. 
The BNL solution $\tilde Z_H(\mu)$ 
and $\tilde\lambda(\mu)$ decrease to zero at the scale 
$\Lambda_{\rm cut}$, as required by the theoretical composite conditions. 
Instead, based on experimental measurements (\ref{thshell}), 
our solution $\tilde Z_H(\mu)\not=0$ up to 
the scale $\Lambda\approx 5.1$ TeV, where the effective quartic coupling 
$\tilde\lambda(\mu)$ becomes negative. It indicates the possible new physics 
(\ref{comp}) and (\ref{bhlx}) discussed in Sec.~\ref{uv}. 
\comment{Note when BHL adopted the compositeness conditions (\ref{bhlc}), the experimental top-quark and Higgs masses were not yet available for using the mass-shell conditions (\ref{thshell}).   
Thus BHL has not considered the intermediate scale $\Lambda$ 
($v<\Lambda\ll \Lambda_{\rm cut}$), where the new physics of 
effective field theory of composite particles $\Phi$ and 
$\Psi$ (\ref{comp}) occurs. 
}

\comment{
It means that the UV completion of our solution is 
drastically different from that of the BHL solution. We give below
a brief summary of our UV completion, more details can 
be found in Refs.~\cite{Xue:2014opa,Xue2017,Leonardi:2018jzn}.
}

We give some discussions how our solutions (Fig.~\ref{figy}) match the effective theory of composite particles (\ref{comp}) at the scale $\Lambda$. 
As the energy scale $\mu$ increases, the four-fermion interaction becomes strong. Its dynamics run away from the weak-coupling asymmetric phase, where the IR ground state for the SM is, and enter the strong-coupling symmetric phase, where the
UV ground state for composite particles is.  
In such a transition from the IR ground state to the UV one, the composite 
Higgs $H\sim \bar t t$ of the SM becomes a more massive and tightly bound state of composite boson $\Phi\sim (\bar t t)$ beyond the SM. 
The $\Phi$ can also combine with $t$ and $b$
quarks to form composite fermions $\Psi$ of $(\bar t t)t$ and 
$(\bar t t)b$. The transition should be described by RG equations for effective couplings and form factors, 
running from the scale $v$ (IR domain) to the scale 
$\Lambda$ (UV domain). 
The BHL effective Lagrangian (\ref{eff}) of the composite Higgs boson for the SM 
is sewed together with the effective Lagrangian of the composite bosons 
$\Phi$ and fermions $\Psi$ by matching at the scale $\Lambda$ 
their form factors $Z_\Phi$ and $Z_\Psi$ to those in Eq.~(\ref{boun0}). These are non-perturbative issues and will be subjects for future numerical studies.

Because the new scale $\Lambda$ is not much larger than the 
electroweak scale $v$. Therefore, it deserves to study the $d>4$ irrelevant operators $\Delta L_{\rm irr}$ (\ref{eff}) suppressed at least $(v/\Lambda)$ 
to find experimentally sizable corrections to the SM observables, for instance,
the $W$ boson mass.     

\section
{\bf $W$ boson right-handed coupling at high energies}\label{ESBSw}

\begin{figure}
\begin{center}
\includegraphics[height=2.00in]{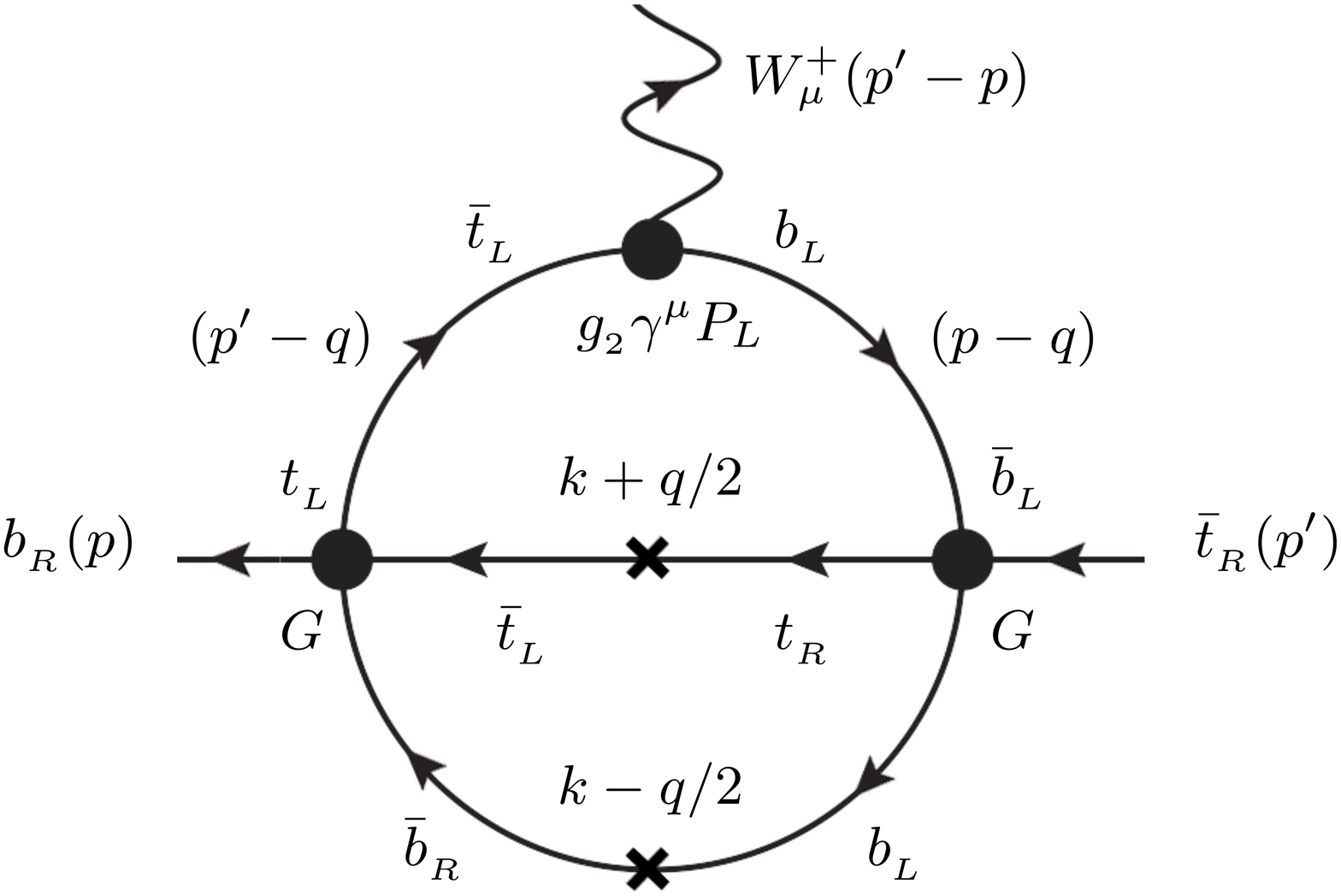}
\caption{We adopt the third quark family $(t,b)$ as an example to illustrate 
the 1PI vertex function of $W$-boson coupling to 
right-handed Dirac fermions induced by four-fermion operators (\ref{bhlx}).
$\ell$, $p$ and $p'$ are external momenta, $q$ and $k$ are internal momenta integrated up to the energy scale $\Lambda$.
The cross ``$\times$'' represents the self-energy functions of Dirac fermions,
which are the eigenstates of a mass operator. 
The $g_2\gamma_\mu P_L$ is the $W$ boson left-handed coupling vertex of the SM. 
This figure and caption are reproduced from figure 3 of Ref.~\cite{Xue:2015wha}}\label{figi}
\end{center}
\end{figure}

Among $\Lambda$-suppressed operators $\Delta L_{\rm irr}$ (\ref{eff}), 
we show \cite{Xue:2015wha} as an example that the  
Feynman diagram of Fig.~\ref{figi} induces 
an effective one-particle-irreducible (1PI) vertex function of $W$-boson right-handed gauge coupling,
\begin{eqnarray}
\Gamma_\mu^W(p',p)&=& i\frac{\bar g_{2}}{\sqrt{2}}\gamma_\mu P_R\,{\mathcal G}^W_R(p',p),\quad q=p'-p,
\label{vr}
\end{eqnarray}
and the dimensionless ${\mathcal G}^W_R(p',p)$ is a Lorentz scalar. 
The external four momenta $q$, $p'$ and $p$ 
are respectively for the $W$ boson, top and bottom quarks. 
Figure \ref{figi} is a complicated two-loop diagram, i.e., a sun-set diagram interacting with an external $W$ boson wave line. 
The ${\mathcal G}^W_R(p',p)$ receives contributions from 
two internal four-momenta integrated up to 
$\Lambda$, $G^2\sim \Lambda^{-4}$ and $G^2\Lambda^4\sim (8\pi^2/N_c)$ \footnote{More detailed calculations can be found there. 
As an example, the left-handed projector 
$P_L$ moves clockwise or anticlockwise to the interacting vertex 
$G$, giving the right-handed operator ${\mathcal G}^W_R(p',p)$. 
Here we do not use the old notation $\Gamma^W(p',p)$ 
to avoid confusion with the $W$-boson decay width.}. From 
the Lorentz invariance and gauge symmetry (the Ward identity),
the 1PI vertex function (\ref{vr}) can be expressed as the 
difference between two sun-set diagrams 
$\Sigma(p')\propto p'^2$ and $\Sigma(p)\propto p^2$, namely 
${\mathcal G}^W_R(p',p)\propto (G\Lambda^2)^2(p^{'2}-p^2)/\Lambda^2$. 
Here, we treat the effective $W$ boson right-handed coupling (\ref{vr}) 
as a model for phenomenological studies.

To find the finite terms at the leading order $(\mu/\Lambda)^\delta$ 
for the least $\delta$, we estimate  in three cases the 
nontrivial ${\mathcal G}^W_R(p',p)$ for given external 
particles' momenta and masses $\mu =p',p, q, m_{b,t},M_W$.
\begin{enumerate}[(i)]
\item  $W$ boson and two fermions are at low energies $\mu=p',p,q \ll\Lambda$, 
\begin{eqnarray}
{\mathcal G}^W_R(p',p)&\sim
&(G\Lambda^2)^2(m^2/\Lambda^2)\ll 1, \quad p',p, q <m,
\label{vr1a}\\
{\mathcal G}^W_R(p',p)&\sim
&(G\Lambda^2)^2(~\mu^2/\Lambda^2)\ll1, \quad p',p,q >m.
\label{vr1b}
\end{eqnarray}
Even for the top and bottom quark case $m^2=m_tm_b$, Eq.~(\ref{vr1a}) 
is negligible for $\Lambda \gtrsim 1$ TeV, and 
thus irrelevant for low energies. We have used Eq.~(\ref{vr1b}) 
as a momentum independent parameter ${\mathcal G}_R$
for studying the right-handed sterile neutrino radiative decay 
\cite{Haghighat:2019rht} and experiment XENON1T \cite{Shakeri2020}.
\item  $W$ boson and one fermion are 
at high energies, while another fermion is at low energies, 
e.g., $\mu=q\approx p'\lesssim \Lambda$ and $p\ll \Lambda$, 
\begin{eqnarray}
{\mathcal G}^W_R(p',p)\sim
(G\Lambda^2)^2(\mu^2/\Lambda^2)\lesssim 1.
\label{vr2}
\end{eqnarray}
It is the case for studying the mass relation of top and bottom 
quarks \cite{Xue:2015wha}.
\item $W$ boson is at low energies $q=p'-p\ll \Lambda$, while 
two fermions are at high energies $p'\approx p \lesssim \Lambda$, 
\begin{eqnarray}
{\mathcal G}^W_R(p',p)\sim
(G\Lambda^2)^2(|q|/\Lambda)< 1.
\label{vr3}
\end{eqnarray}
It gives the contribution to the effective low-energy 
$W$-boson coupling $\bar g_2(q)$, if two fermions $p'$ and $p$ 
are internal momenta of the bubble diagrams for the 
$W$-boson vacuum polarisation tensor.  
\end{enumerate}
The effective right-handed vertex function ${\mathcal G}^W_R(p',p)$ has the following properties. For low energies $p',p,q\ll \Lambda$, ${\mathcal G}^W_R(p',p)\rightarrow 0$ recovering the SM parity-violating 
symmetry. While ${\mathcal G}^W_R(p',p)\rightarrow {\mathcal O}(1)$ for high energies $p',p,q\lesssim \Lambda$, implying 
the parity-symmetry restoration at high energies \cite{xue2013}. It is
consistent with the vectorlike spectra of composite particles (\ref{comp}).

As a result, the $W$-boson coupling is no longer purely left-handed. 
We generally assume that the charged 
$W$-boson $SU_L(2)$ coupling $\bar g_2$ takes the form \footnote{We obtained the effective interaction (\ref{vrw}) because the chiral gauge symmetry can not exactly preserve in any regularised quantum field theory \cite{Xue1997, Xue1999}.}.
\begin{eqnarray}
i\frac{\bar g_{2}}{\sqrt{2}}\gamma_\mu \big[V_LP_L+V_RP_R\,{\mathcal G}^W_R(p',p)\big].
\label{vrw}
\end{eqnarray}
Since all SM fermions 
process four-fermions interactions 
$G\psi_L\psi_R\psi_R\psi_L$ \cite{Xue:2016dpl,Xue2017} in the same structure as 
the top and bottom channel (\ref{bhlx}), the interaction (\ref{vrw}) 
is generalised to all families of SM leptons and quarks. 
The family mixing matrix $V_L$ is the CKM matrix ${\mathcal U}_L^{u\dagger}{\mathcal U}_L^d$ for quarks 
or the PMNS matrix ${\mathcal U}_L^{\nu\dagger}{\mathcal U}_L^\ell$ 
for leptons. Whereas, their 
counterparts $V_R$ in the right-handed sector of ${\mathcal U}_R^{u\dagger}{\mathcal U}_R^d$ for quarks or 
the ${\mathcal U}_R^{\nu\dagger}{\mathcal U}_R^\ell$ 
for leptons \cite{Xue:2016dpl}. 
In family space, all unitary matrices ${\mathcal U}_{L, R}$ transform quark and lepton 
fields from gauge eigenstates to mass eigenstates. 

We generalize the above discussions to the neutral gauge boson channels:   
the $SU_L(2)$ coupling $\bar g_2t^i_3\bar\psi^i_L\gamma^\mu \psi^i_L W_\mu^3$, the $U_Y(1)$ coupling $\bar g_1t^i_3\bar\psi^i_L\gamma^\mu \psi^i_L B_\mu$ and $\bar g_1q_i\bar\psi^i\gamma^\mu\psi^i B_\mu$. Here $t^i_3=(1/2,-1/2)$ is the third isospin component, $q_i$ is the electric charge, and 
the isospin index $i=t,b$ for the top and bottom quarks. In the  
Feynman diagram \ref{figi}, replacing $t$ ($b$) by $b$ ($t$) and 
$W^+_\mu$ by $W^3_\mu$ or $B_\mu$, we obtain the induced 1PI vertex functions of $W^3_\mu$ and $B_\mu$ right-handed gauge couplings. 
The neutral $SU_L(2)$ field $W^3_\mu$  and $U_Y(1)$ field $B_\mu$  
vertexes coupling to the left-handed fermions are modified by
\begin{eqnarray}
i\frac{\bar g_{2}}{\sqrt{2}}t^i_3\gamma_\mu \big[P_L+ P_R\,{\mathcal G}^W_R(p',p)\big];\quad  i\frac{\bar g_{1}}{\sqrt{2}}t^i_3\gamma_\mu \big[P_L+ P_R\,{\mathcal G}^Z_R(p',p)\big].
\label{vrwn}
\end{eqnarray}
The 1PI function ${\mathcal G}^Z_R(p',p)$ has the same properties 
(i), (ii) and (iii) of Eqs.(\ref{vr1a}-\ref{vr3}). 
The family mixing matrices $V_L$ and $V_R$ are absent in modified 
vertexes (\ref{vrwn}). 
\comment{We assume 
${\mathcal G}^W_R(p',p)\sim {\mathcal G}^Z_R(p',p)$ from the 
view point of parity-symmetry restoration and chiral gaueg couplings 
$\bar g_{2}$ and $\bar g_{1}$ becoming vector-like at the scale $\Lambda$.
}

To end this section, we briefly compare and contrast our model with the manifest left-right symmetry (LR) model \cite{Beg1977, Chakrabortty2012}. 
The similarity between the two models is the parity-symmetry restoration at high energies. The most evident
differences are in gauge symmetries and bosons.
The LR model introduces the new gauge symmetry $SU_R(2)$ and extra gauge 
bosons, e.g.~$W_R$, as well as an additional Higgs sector for the
$SU_R(2)$ symmetry breaking and 
gauge bosons $W_R$ masses generation. While our model possesses only the SM gauge symmetries and gauge bosons $W$ and $Z$.
 
\section{$W$ and $Z$ boson mass corrections}\label{wzmass}

Using the modified couplings (\ref{vrw}) and 
(\ref{vrwn}), we compute the radiative corrections (bubble diagrams) to 
the $W$ and $Z$ boson propagators and find the right-handed corrections 
to their masses, namely the corrections from the right-handed 
couplings ${\mathcal G}^{W,Z}_R(p',p)$. For the (iii) case (\ref{vr3}) of 
low energy $W$ and $Z$ bosons $q=p'-p\approx M_{W,Z}\ll \Lambda$, while 
two fermions are at high energies $p'\approx p \lesssim \Lambda$, 
we parametrise the right-handed 1PI functions as 
\begin{eqnarray}
{\mathcal G}^W_R(q)=c_w|q|/\Lambda;\quad  {\mathcal G}^Z_R(q) = c_z|q|/\Lambda ,
\label{vrwg}
\end{eqnarray}
where the scale $\Lambda\approx 5.1$ TeV and coefficients 
$c_w$ and $c_z$ should be the order of unity ${\mathcal O}(1)$. 
Here the scale $\Lambda$ reflects the characteristic scale for 
the parity symmetry restoration. Namely, when $q\rightarrow \Lambda$
the $W$ and $Z$ boson couplings tend to become vectorlike. 

In the top and bottom quark channel, following the 
BHL approach \cite{Bardeen1990} and neglecting the bottom quark mass $m_b$, 
we arrive at the renormalised and gauge-invariant $W$ and $Z$ propagators
\begin{eqnarray}
D_{\mu\nu}^W(q) &=&\frac{q_\mu q_\nu/q^2-g_{\mu\nu}}{q^2- \bar g_{2}^2(q)\big[1+c_w^2(|q|/\Lambda)^2\big]\bar f^2(q)},
\label{Wpro}\\
D_{\mu\nu}^Z(q) &=&\frac{q_\mu q_\nu/q^2-g_{\mu\nu}}{q^2- \big[\bar g^2_1(q)+\bar g^2_2(q)\big]\big[1+c_z^2(1/2)(|q|/\Lambda)^2\big]f^2(q)}. 
\label{Zpro}
\end{eqnarray}
The $\bar f^2(q)$ and $f^2(q)$ are the decay constants of the 
charged and neutral Goldstone bosons, which become the longitudinal modes 
of massive $W$ and $Z$ bosons. The energy running $W^\pm$ and $Z^0$ masses are  
\begin{eqnarray}
M_W^2(q)&=& \bar g_{2}^2(q)\big[1+c_w^2(|q|/\Lambda)^2\big]\bar f^2(q),
\label{Wmassq}\\
M_Z^2(q)&=& \Big[\bar g_1^2(q)+\bar g_2^2(q)\Big]\big[1+2^{-1}c_z^2(|q|/\Lambda)^2\big]
f^2(q),
\label{Zmassq}
\end{eqnarray}
and the energy scale $q^2=\mu^2$. These are similar to the running 
top-quark mass $m_t(\mu)=\bar g_t^2(\mu)v/\sqrt{2}$ and Higgs 
mass $m_{_H}(\mu)=[2\bar \lambda(\mu)]^{1/2} v$, see Eq.~(\ref{thshell}).
At the tree-level SM Lagrangian, the right-handed couplings 
(\ref{vrwg}) vanish, the gauge couplings are
$\bar g_{1,2}^2(0)$ and  
the SM gauge-boson masses $M_W=(1/2)\bar g_{2}(0)v$ and 
$M_Z=(1/2)(g_2^2(0)+g_1^2(0))^{1/2}v$ in terms of 
the electroweak scale $v=(\sqrt{2}G_F)^{-1/2}\approx 246 {\rm GeV}$. 

\begin{figure*}[t]
\begin{center}
\includegraphics[height=6.5cm,width=8.8cm]{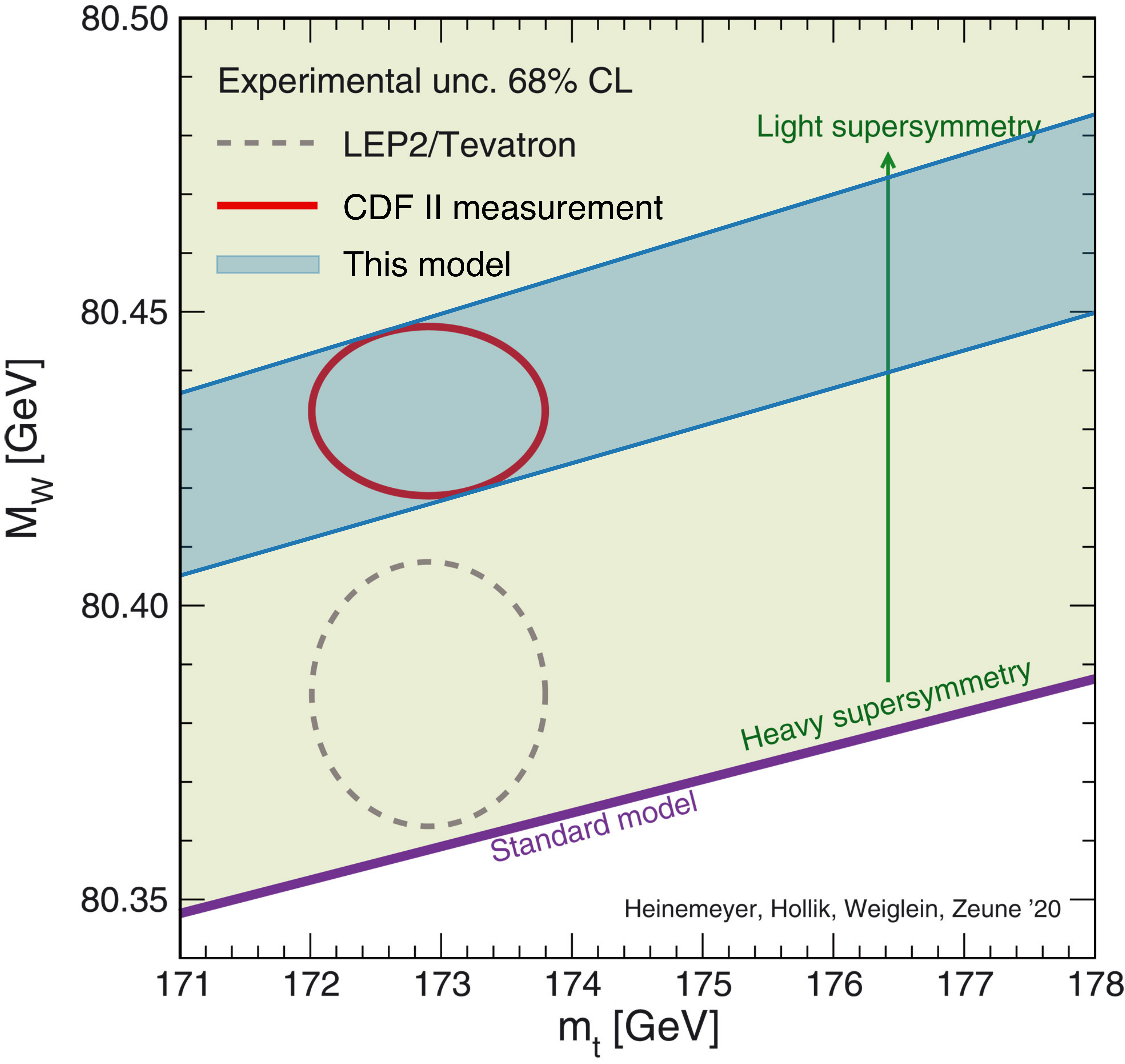}
\caption{This figure is a modified version of 
Fig.~1 of Refs.~\cite{Heinemeyer2013,Aaltonen2022}, and the detailed captions can be 
found there. The  ``Standard Model'' line indicates the SM result with the Higgs mass $M_H^{\rm SM}=125.6\pm 0.7$ GeV. The green shaded region shows the result from supersymmetric extensions to the SM. 
The blue shaded region represents the result (\ref{Wmassc}) of the scale 
$\Lambda\approx 5.1$ TeV and the upper (lower) bound 
$c_w\approx 2.09$ $(1.68)$ from the right-handed corrections.}
\label{figwm}
\end{center}
\end{figure*}

In the absence of the right-handed 
correction $c^2_w(M_W/\Lambda)^2$, 
the top-quark mass $m_t=\bar g_t(m_t)v/\sqrt{2}$ and $W$-boson mass
$M_W=\bar g_{2}(M_W)v/2$ yield 
\begin{eqnarray}
M_W=\big[\bar g_{2}(M_W)/\bar g_t(m_t)\big]m_t/\sqrt{2},
\label{Wtmass}
\end{eqnarray}
where the $W$-boson gauge coupling $\bar g_{2}(M_W)$ is the RG solution 
in the SM, and the Yukawa coupling $\bar g_t(m_t)$ is the RG 
solution (Fig.~\ref{figy}) obtained by the experimentally measured  
top-quark and Higgs masses (\ref{thshell}).
From Eqs.~(\ref{Wmassq}) and (\ref{Zmassq}), we define
the experimentally measured $W^\pm$ and $Z^0$ masses on 
their mass-shell conditions,
\begin{eqnarray}
M^{\rm exp}_W&=& \bar g_{2}(q)\big[1+c_w^2(|q|/\Lambda)^2\big]^{1/2}\bar f(q)\Big|_{q^2=M_W^2}\nonumber\\
&\approx & M^{\rm SM}_W\big[1+c_w^2(M^{\rm SM}_W/\Lambda)^2\big]^{1/2},
\label{Wmassc}\\
M^{\rm exp}_Z&=&[\bar g_1^2(q)+\bar g_2^2(q)]^{1/2}\big[1+2^{-1}c_z^2(|q|/\Lambda)^2\big]^{1/2}f(q)\big|_{q^2=M_Z^2}\nonumber\\
&\approx & M^{\rm SM}_Z\big[1+2^{-1}c_z^2(M^{\rm SM}_Z/\Lambda)^2\big]^{1/2},
\label{Zmassz}
\end{eqnarray}
up to the leading order of the correction $c^2_w(M^{\rm SM}_W/\Lambda)^2$ or $c^2_z(M^{\rm SM}_Z/\Lambda)^2$. 
The $M^{\rm SM}_W$ and $M^{\rm SM}_Z$ stand for 
the gauge boson masses that 
receive the full high-order corrections in the SM with the measured 
Higgs mass $m_{_H}=126$ GeV \cite{Heinemeyer2013}.   
Using the relation (\ref{Wtmass}), we plot in Fig.~\ref{figwm} the
corrected $W$-boson mass (\ref{Wmassc}) in the blue area 
for the values $1.68\leq c_w\leq 2.09$. 
It shows that for these $c_w$ values of the order of unity, 
the result is consistent with the recent 
experimental result $M^{\rm exp}_W$ (the red elliptical circle) 
for $m_t\approx 173$ GeV. 
\comment{Thus the $W$-boson mass tension can be relieved by the correction from the $W$-boson right-handed gauge 
coupling (\ref{vrwg}). 
Using the relation (\ref{Wtmass}), we plot in Fig.~\ref{figwm} the
corrected $W$-boson mass (\ref{Wmassc}) for selected 
$\Lambda$ values in the range of $2.44$ and $3.03$ TeV. 
It shows that for these 
$\Lambda$ values the result is consistent with the recent 
experimental result $M^{\rm exp}_W$ (the red elliptical circle) 
for $m_t\approx 173$ GeV. Thus the $W$-boson mass tension can be relieved by the correction from the $W$-boson right-handed gauge 
coupling (\ref{vrwg}). 
It is remarkable that the $\Lambda$ values constrained by the $W$-boson 
mass $M^{\rm exp}_W$ are consistently in the same order of magnitude with the scale $\Lambda\approx 5$ TeV (Fig.\ref{figy}) of the new physics beyond SM. That is independently obtained from the solutions to the 
low-energy RG equations (\ref{reg1},\ref{reg2}) and experimental top-quark and Higgs masses (\ref{thshell}). The result of Fig.~\ref{figwm}  
implies that the proposed theoretical scenario beyond SM is 
self-consistent with the new physics appearing in TeV scales.
}

The $Z$-boson mass (\ref{Zmassz}) also receives the right-handed correction. 
The constraint $c_z<3.79 \times 10^{-1}$ is given by 
the uncertainty in the reported measurement 
$M^{\exp}_Z=91.1876\pm 0.0021$ GeV \cite{Tanabashi2018}. The correction to 
the on-shell $\cos^2\theta_W$ reads
\begin{eqnarray}
\cos^2\theta_W\equiv \frac{M_W^2}{M_Z^2} \approx \cos^2\theta_W\Big|_{\rm SM} \Big[1+ (c_w^2-2^{-1}c_z^2)(|q|/\Lambda)^2\Big]_{|q|=M_W} .
\label{cos}
\end{eqnarray}
The correction $\delta \cos^2\theta_W \lesssim 5\times 10^{-4}$ 
is consistent to the experimental uncertainty on the 
$\delta \cos^2\theta_W=-\delta \sin^2\theta_W < 2\times 10^{-3}$ \cite{Bentz2010,Tanabashi2018}. The parameter $\rho=M^2_W[M^2_Z\cos^2\theta_W]^{-1}=1$ does not receive the right-handed correction. 
\comment{
$\sin^2\theta_W(q)$The running masses (\ref{Wmassq}) $M_W(q)$ and $M_Z(q)$ (\ref{Zmassq}) receive the same corrections. Thus
no corrections give to the $\cos^2\theta_W(q)=M^2_W(q)/M^2_Z(q)$, 
$\sin^2\theta_W(q)$ and $\rho=[M^2_W(q)/M^2_Z(q)]\cos^{-2}\theta_W(q)]=1$. 
On the mass shells (\ref{Wmassc}) and (\ref{Zmassz}), 
$\cos^2\theta_W=M^2_W(M_W)/M^2_Z(M_Z)$ receives a correction 
$(M_W^2-M_Z^2)/\Lambda^2 \lesssim {\mathcal O}(10^{-4})$. This is the reason for which we implement the approximation ${\mathcal G}^Z_R(q)\approx {\mathcal G}^W_R(q)$. As a result, unlike the modified 
$W$-boson coupling (\ref{vrw}), there are not 
the leading-order ${\mathcal G}^Z_R(q)$ and ${\mathcal G}^W_R(q)$ 
corrections to the $Z_0$ boson gauge coupling to fermions in SM.
}     

\section{$W$ and $Z$ boson width corrections}\label{wzwidth}
We turn to examine the right-handed corrections to the $W$ and $Z$ boson 
decay widths $\Gamma_{W,Z}$. It will be a crosscheck of the present 
theoretical scenario. Analogously to the masses $M_{W,Z}$, the widths 
$\Gamma_{W,Z}$ are also the SM quantities fixed without 
any free parameter after the Higgs mass is measured. Calculating total width 
of $W$-boson decaying to all leptons and quarks, except $t$ and $b$ quarks,
see for example Refs.\cite{Donoghue2014,Halzen1984a}, we obtain in the SM
\begin{eqnarray}
\Gamma^{\rm SM}_W=\frac{3}{16\pi}\bar g^2_{2}\big|^{\rm SM}M_W^{\rm SM}
\approx 2.07 ~ {\rm GeV},
\label{smw}
\end{eqnarray} 
using the unitary conditions $\sum_{\ell=e,\mu,\tau}|V_L^{i\ell}|^2=1$ 
for $i=\nu_e,\nu_\mu,\nu_\tau$, and $\sum_{q=d,s,b}|V_L^{iq}|^2=1$ 
for $i=u,c$. The decay to the top and bottom quark channel ($i=t$) is 
kinetically forbidden. The superscript ``SM'' stands for full SM high-order corrections have been taken into account. In the same way as the SM calculations, considering the modified $W$-boson coupling (\ref{vrw}) 
we obtain up to the leading order
\begin{eqnarray}
\Gamma^{\rm exp}_W &\approx& \frac{3}{16\pi}\bar g^2_{2}\big|^{\rm SM}
\big[1+({\mathcal G}^W_R)^2\big]^{1/2}
M_W^{\rm SM}\Big[1+c_w^2(M^{\rm SM}_W/\Lambda)^2\Big]^{1/2}\nonumber\\
&\approx& \Gamma^{\rm SM}_W\Big[1+c_w^2(M^{\rm SM}_W/\Lambda)^2\Big]^{1/2},
\label{smwr}
\end{eqnarray} 
using the unitary conditions $\sum_{\ell=e,\mu,\tau}|V_R^{i\ell}|^2=1$ 
for $i=\nu_e,\nu_\mu,\nu_\tau$ and $\sum_{q=d,s,b}|V_R^{iq}|^2=1$ for $i=u,c$. 
Right-handed corrections come from two aspects. One is the mass 
$M^{\rm SM}_W$ correction (\ref{Wmassc}). Another is the correction factor
$\big[1+({\mathcal G}^W_R)^2\big]^{1/2}$ to the $W$-boson coupling 
$\bar g^2_{2c}\big|^{\rm SM}$ vertex accounting for the decay. 
In the decay process, the $W$ boson and two fermions are at low energies $\mu\sim M_W\ll \Lambda$, the right-handed coupling ${\mathcal G}^W_R\sim (M^{\rm SM}_W/\Lambda)^2$, as discussed in the (i) case 
(\ref{vr1b}). Therefore, at the leading order correction 
${\mathcal O}[(M^{\rm SM}_W/\Lambda)^2]$,
the $W$-boson decay width receives the same correction (\ref{smwr}) as its mass (\ref{Wmassc}).   
Based on the experimental measurement 
$\Gamma_W^{\rm exp}=2.085\pm 0.042$ GeV \cite{Tanabashi2018}, 
and the values $1.68\leq c_w\leq 2.09$ (\ref{Wmassc}) fixed by
the high-precision $W$ mass measurement \cite{Aaltonen2022}, we obtain 
\begin{eqnarray}
\Gamma_W\approx \Gamma^{\rm SM}_W[1 + (3.51\sim 5.44)\times 10^{-4}].
\label{smwc}
\end{eqnarray}
Using the experimental $\Gamma_W^{\rm exp}$ centre value $2.085$ GeV for 
$\Gamma^{\rm SM}_W$, 
we find the right-handed correction is within the error bar $\pm 0.042$ GeV. 

In the same way, considering the 
modified $Z$-boson coupling (\ref{vrwn}), we obtain the leading order 
right-handed correction to the $Z$ boson decay width 
\begin{eqnarray}
\Gamma^{\rm exp}_Z &\approx& \Gamma^{\rm SM}_Z\Big[1+c_z^2(M^{\rm SM}_Z/\Lambda)^2\Big]^{1/2},
\label{smzr}
\end{eqnarray}
which is the same as the $Z$ boson mass correction. 
Based on the experimental measurement 
$\Gamma_Z^{\rm exp}=2.4952\pm 0.0023$ GeV \cite{Tanabashi2018} and the 
constraint $c_z < 3.79\times 10^{-1}$ (\ref{Zmassz}) from the uncertainty in
the $M_Z$ measurement, the positive correction to the $Z$ boson width (\ref{smzr}) should be smaller than 
$5.73\times 10^{-5}$ GeV. It is within the error bar $\pm 0.0023$ GeV of the measurement $\Gamma_Z^{\rm exp}$. 
Therefore, the high-precision measurements of the $Z$ boson mass, 
the $W$ and $Z$ boson decay widths are important for cross-checks. 

\section{Conclusion and remarks
}
\hskip0.1cm
We discuss the possible new physics scenario beyond the SM.
One of the features is that the $W$ and $Z$ boson gauge couplings are not 
maximally parity-violating, non-vanishing right-handed couplings depend on momenta, and the parity symmetry could restore
around the energy scale $\Lambda\sim$ TeV scales. We adopt the value 
$\Lambda\approx 5.1$ TeV indicated by the composite Higgs high-energy 
behaviours, which are solutions to the RG equations demanding mass-shell 
conditions of experimentally measured v.e.v $v$, top-quark and Higgs masses. 
The $W$-boson right-handed coupling
gives rise to the leading order right-handed correction 
$c_w^2(M^{\rm SM}_W/\Lambda)^2$ to the $W$ boson mass of the SM. 
For $c_w^2\sim {\mathcal O}(1)$ be the order of unity, 
this correction consistently 
relieves the $W$-boson mass $7 \sigma$ 
tension recently discovered \cite{Tanabashi2018}. 
The analogous corrections contribute to the $Z$ boson mass, the $W$ and $Z$ 
boson decay widths. They are tiny and within the error bars of existing 
experimental measurements. The high-precision measurements of 
these quantities will be important for cross-checks. 

We mention that the asymmetry measurements can possibly examine the right-handed correction to the $W$ boson gauge coupling.  
In general, it was suggested \cite{Xue2003,xue2013} to measure the asymmetry 
\begin{eqnarray}
A_{L,R}&=&\frac{\sigma_L-\sigma_R}{\sigma_L+\sigma_R}\approx 1-2[{\mathcal G}^W_R(\mu)]^2,
\label{as0}
\end{eqnarray} 
and $\sigma_{L,R}$ are cross-sections of left- and right-handed 
polarised particles in interactions. Which energy-dependent case 
of (\ref{vr1a}-\ref{vr3}) for the right-handed coupling 
${\mathcal G}^W_R(\mu)$ depends on the 
process under consideration. 
It is expected that $\sigma_R= [{\mathcal G}^W_R(\mu)]^2\sigma_L\rightarrow 0$ 
and $A_{L,R}\rightarrow 1$ for $\mu\ll \Lambda$, whereas $A_{L,R}\rightarrow 
0$ for $\mu\sim \Lambda$. The asymmetry signatures should be more evident in 
high energies, for example the CDF \cite{Aaltonen2008} and 
D0 \cite{Abazov2008} experiments that measured the forward-backwards asymmetry 
in top-quark pair production at the Fermilab Tevatron $p\bar p$ collisions.



\providecommand{\href}[2]{#2}\begingroup\raggedright\endgroup

\end{document}